\documentclass[3p,times,twocolumn]{elsarticle}

\usepackage{amssymb}

\usepackage[hyperfootnotes=false]{hyperref}

\newcommand\blfootnote[1]{%
  \begingroup
  \renewcommand\thefootnote{}\footnote{#1}%
  \addtocounter{footnote}{-1}%
  \endgroup
}

\begin{document}

\begin{frontmatter}




\title{Ultra-high energy neutrino searches and GW follow-up with the Pierre
Auger Observatory}


\author{M.~Schimp$^\mathrm{a}$ for the Pierre Auger Collaboration$^\mathrm{b}$}

\address{$^a$Bergische Universit\"at Wuppertal, Gau\ss{}str. 20, D-42119, Wuppertal\\
$^b$Av. San Martin Norte 304, 5613 Malarg\"ue, Argentina\\ (Full author list: \url{http://www.auger.org/archive/authors\_2018\_06.html})}

\begin{abstract}
The surface detector array (SD) of the Pierre Auger Observatory is
sensitive to neutrinos at energies in the 100~PeV to 100~EeV range.
This sensitivity, together with its large acceptance, makes it a complementary detector to other
neutrino telescopes, which have their peak sensitivities at lower energies.
The neutrino-induced air showers that the SD of the Pierre Auger
Observatory is sensitive to can be divided into those induced by
interactions of neutrinos of any flavor deep in the atmosphere,
and those induced by charged-current interactions of tau neutrinos in the Earth's crust.
Both of these types can be efficiently distinguished from cosmic ray-induced air showers, provided that their
zenith angles are larger than 60$^\circ$.
As no neutrino candidates were found in the performed searches, we
present limits on the diffuse all-flavor neutrino flux.
Using these limits, we obtained constraints on cosmic-ray and
neutrino production models.
In the light of the recent observations of gravitational waves (GW), we
also present the follow-up of LIGO/Virgo GW events.
These include binary black hole merger events and also GW170817,
the only binary neutron star merger ever observed directly.
\end{abstract}


\end{frontmatter}


\section{Introduction}
\label{sec:intro}
\blfootnote{\textcopyright{} 2019. Licensed under terms of the Creative Commons Attribution-NonCommercial-NoDerivatives 4.0 International License \href{https://creativecommons.org/licenses/by-nc-nd/4.0/}{(CC-BY-NC-ND-4.0)}}
The search for the origin of ultra-high energy cosmic rays (UHECRs) has been going on for many decades and a final answer is still to be found. One piece of the puzzle of understanding the origin of UHECRs could be ultra-high energy neutrinos with energies above $0.1~\mathrm{EeV}$ (UHE neutrinos).
Despite UHE neutrinos have not been found yet, they are thought to originate from UHECRs, being produced either at the sources of UHECRs or during their propagation through the Universe~\cite{greisen1966end, zatsepin1966upper, hooper2005impact, ave2005cosmogenic, kotera2010cosmogenic}.
Neutrinos have the useful property to be, on the one hand, electrically neutral and thus not affected by magnetic fields.
Therefore, they travel in straight lines and point back to their sources.
On the other hand, they are only subject to the weak interaction, leading to almost no attenuation and therefore to practically no neutrino horizon as opposed to photons.

The Pierre Auger Observatory, described in detail in~\cite{NIM}, located close to Malarg\"ue in the Province of Mendoza, Argentina, is, by surface area, the largest operational cosmic-ray detector in the world.
It consists mainly of a surface detector array (SD) and a fluorescence detector (FD).
The SD is a triangular grid of 1660 water-Cherenkov detectors, also called stations, each filled with very pure water, separated by $1.5~\mathrm{km}$ and distributed over an area of $3000~\mathrm{km}^2$.
The FD consists of 27 telescopes, distributed over four sites surrounding the SD.
They record the fluorescence light emitted by the nitrogen in the atmosphere after cosmic-ray induced air showers deposit their energy in it.
The SD is located at an average altitude of 1400~m~a.s.l., corresponding to a vertical atmospheric depth of 875~g/cm$^{2}$.
It has been taking data since 2004 while it was still under construction.

The SD of the Pierre Auger Observatory can be used to search for both, down-going (DG) UHE neutrino induced air showers, as well as air showers from so called Earth-skimming (ES) $\tau$ neutrinos~\cite{aab2015improved}.
The DG showers are classified as either high-zenith ($75^\circ<\theta<90^\circ$; DGH) or low-zenith ($60^\circ<\theta<75^\circ$; DGL) showers, for which the reconstructions and selections are different.
The event characteristics, the search procedure, the results and their implications are described in detail in Section~\ref{sec:searches}.

Having established the search procedure, it is additionally possible to use the UHE neutrino search with the Pierre Auger Observatory as an element in a multimessenger search by following up other astrophysical observations.
This is mainly motivated by the mentioned paradigm that UHE neutrinos are secondary particles, produced in the sources or during the propagation of UHECRs.
One of the newest ways to potentially observe UHECR sources and therefore a new contributor to multimessenger astronomy relevant to UHE neutrino searches is gravitational wave (GW) astronomy which was initiated by the detection of GWs by binary black hole (BBH) mergers in 2015~\cite{abbott2016observation, abbott2016gw151226}.
Therefore, the Pierre Auger Collaboration and the LIGO/Virgo Collaboration (LVC) signed a memorandum of understanding (MoU), allowing the Pierre Auger Collaboration to perform follow-up searches of the GW events detected by LVC.
The procedure and the results of these GW follow-up searches are reported in Section~\ref{sec:GW}.

\section{Searches for ultra-high energy neutrinos with the Pierre Auger Observatory}
\label{sec:searches}
\subsection{Search procedure}
Atmospheric air showers produced by inclined UHE neutrinos (i.e. both ES and DG) that interact deep
in the atmosphere can be distinguished from cosmic-ray induced showers, which have their primary interactions high in the atmosphere, by the different time structure of the signals in the water-Cherenkov detectors.
More specifically, inclined UHE neutrino induced showers starting deep in the atmosphere have a substantial electromagnetic and/or hadronic component that reaches the detectors and produces long-lasting light signals, while cosmic-ray induced showers of the same inclinations have lost these components during their propagation through the atmosphere and consist mostly of muons when they reach the ground, producing much shorter light signals in the detectors.
For zenith angles below~$60^\circ$, the slant depth of the atmosphere is so low that cosmic-ray induced showers often reach the ground with such substantial hadronic and electromagnetic components that they are not sufficiently distinguishable from neutrino induced showers. 
This is the reason for the choice of showers with zenith angles above $60^\circ$.

For each of the channels (DGL, DGH, ES), there are different procedures for the event selection, neutrino identification, and exposure calculation.
These are described in more detail in~\cite{aab2015improved}.
To estimate the efficiency of the measurement in dependence of the search parameters, neutrino events are simulated for various energies, geometries, weak interaction channels (charged and neutral current), interaction depths, and, for the case of ES $\tau$ neutrinos, altitudes of the $\tau$ lepton decay, to sample the physical parameter space.
All searches include a pre-selection of the correct shower inclinations, and a selection based on a variable that is optimized to distinguish between neutrino and cosmic-ray induced events.
A `cut value' for the neutrino selection is defined by comparing the distribution of this variable for simulated neutrino events to that of an unbiased subsample of the data, which represents the background distribution of cosmic-ray induced showers.

For the ES channel, at least three triggered stations are required, and the shape of the area of triggered stations on the ground, the so called footprint, is required to be substantially elongated.
Also, the apparent average speed of the trigger signals at the ground between pairs of stations is required to be close to the speed of light with only a small variance.
For the DGH channel, at least four triggered stations are required.
Additionally, similar criteria as for the ES channel apply but with other ranges for the selection variables, accounting for the lower inclination.
Also, a plane fit of the shower front must yield a zenith angle of at least 75$^\circ$.
The search in the DGL channel relies on standard directional reconstructions like a fit of a curved shower front to the data.
For this channel, the required number of triggered stations in the central hexagon is 4 (5) for the zenith angle range of $67.5^\circ - 75^\circ$ ($60^\circ - 67.5^\circ$).
Having reconstructed the geometry this way, at least 75\% of the triggered stations in the central hexagon of the event are required to fulfill the criteria of the SD's time-over-threshold (ToT) trigger~\cite{NIM} in order to remove a large amount of muon-rich showers, which, due to their narrow time traces, rarely fulfill the ToT trigger criteria and, as discussed earlier, are associated with cosmic-ray induced events.

The key parameter for the construction of the variables that separate neutrinos from cosmic-ray showers is the area over peak (AoP) of the signal in each water-Cherenkov detector.
It is defined as the time-integrated signal divided by the maximum of the signal in each detector, both normalized to corresponding values obtained by calibration to vertically through-going muons.
For the ES search, the variable of choice is simply the average AoP of the triggered stations ($<$AoP$>$).
For the DGH (DGL) channel, a Fisher discriminant approach is used to combine up to 10 variables using the AoPs of 4 (4 or 5, in the angular ranges as described before) early (central) stations~\cite{aab2015improved}.
The background distributions of $<$AoP$>$ and the Fisher discriminant variable, estimated from the data, are found to have exponential tails.
Therefore, the cut value is obtained for each channel by extrapolating the exponential tail of its background distribution and choosing the cut value such that a certain rate of false neutrino candidates from the background distribution is expected.
For the ES and DGH (DGL) channels, this rate is one expected false positive event every 50 (20) years of the SD running with design sensitivity.
In the DGH (DGL) channel, in order to account for differences in the shower size (shower inclination), a subdivision of events is made based on the number of triggered SD stations (zenith angle).
The variable optimizations and estimations of the cut values are done for these subdivisions individually.

\begin{figure}[t]\includegraphics[width=.475\textwidth]{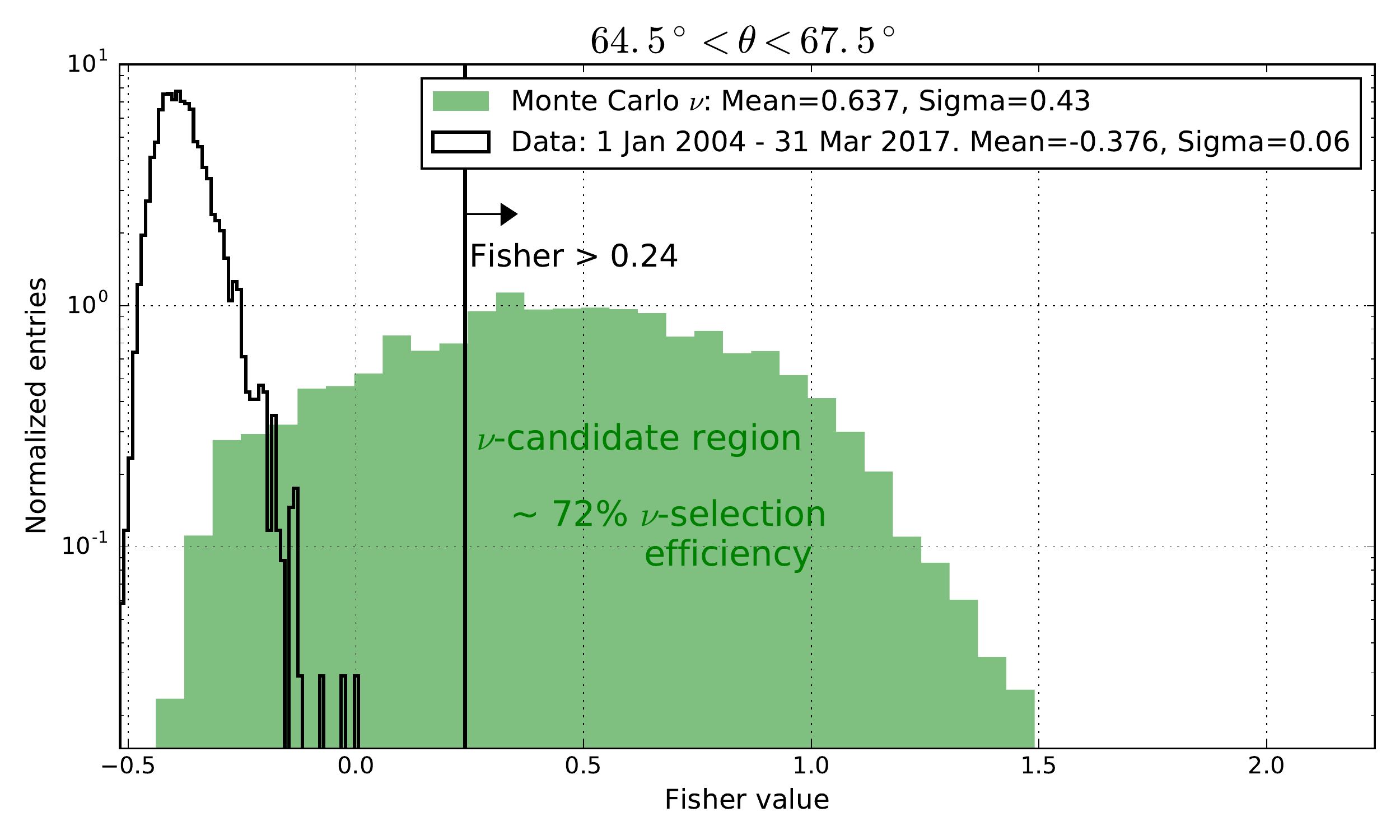}\caption{Fisher distributions for inclinations of $(66\pm1.5)^\circ$ for background estimated from data (s. text; black) and simulated neutrino events (filled green).}\label{fig:FisherDist_66}\end{figure}

\begin{figure}[t]\includegraphics[width=.475\textwidth]{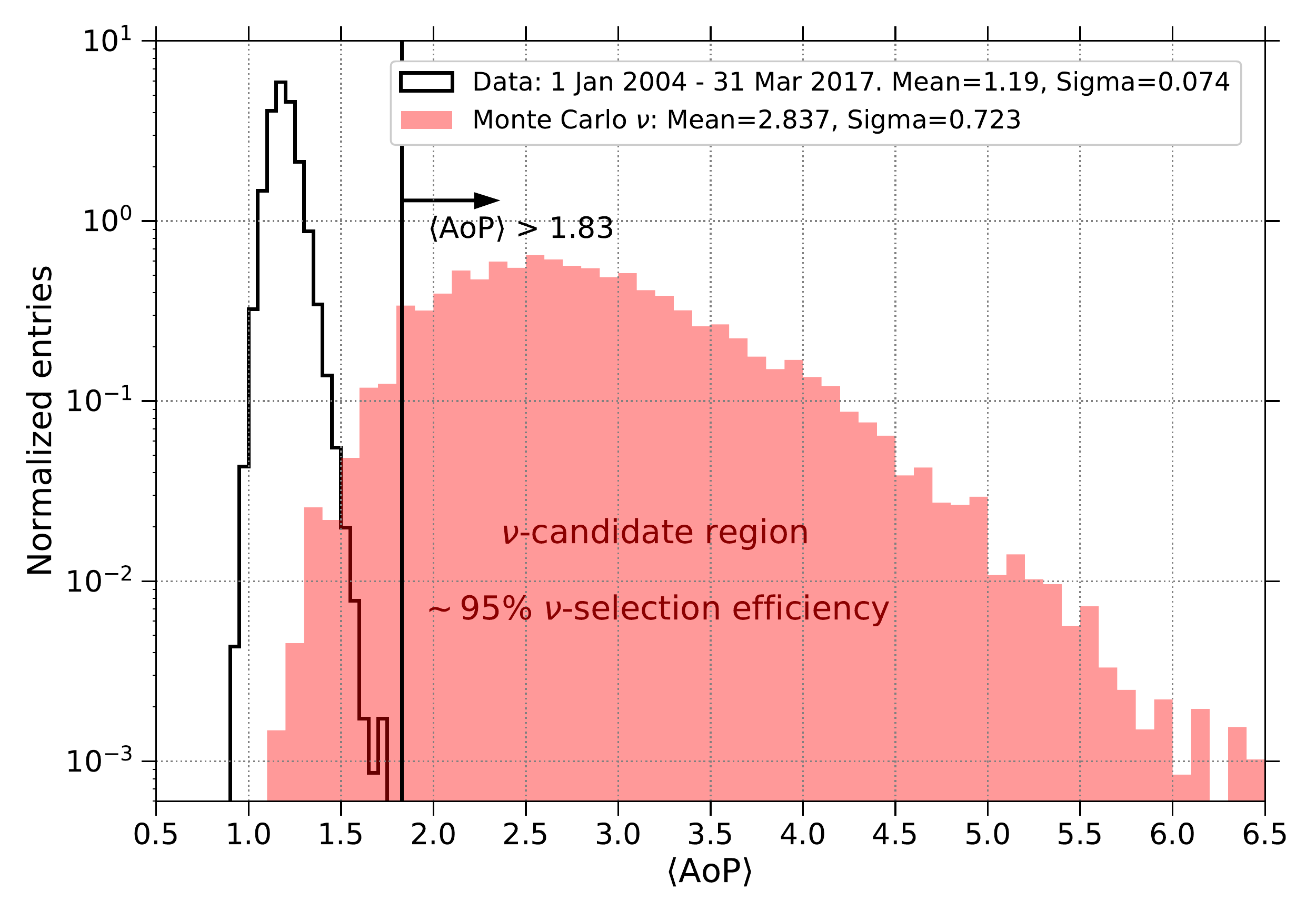}\caption{$<$AoP$>$ distributions for inclinations of $>90^\circ$ for background estimated from data (s. text; black) and simulated neutrino events (filled red).}\label{fig:FisherDist_92.5}\end{figure}

For this work, the data from the period of 2004-01-01 through 2017-03-31 has been used to search for UHE neutrinos and contains no neutrino candidates~\cite{zas2017searches}.
In Figures~\ref{fig:FisherDist_66} and~\ref{fig:FisherDist_92.5}, exemplary distributions of the critical variables are shown for the simulated neutrino-induced showers and the measured data, for two different angular regions.
The exponential tails of the estimated background distributions are clearly visible.
I can also be seen that the search is more efficient at higher inclinations, which is due to the better separation between signal and background.

\subsection{Limits on the diffuse flux of UHE neutrinos and on point-like UHE neutrino sources}
\label{subsec:limits}
\begin{figure}[t]\includegraphics[width=.475\textwidth]{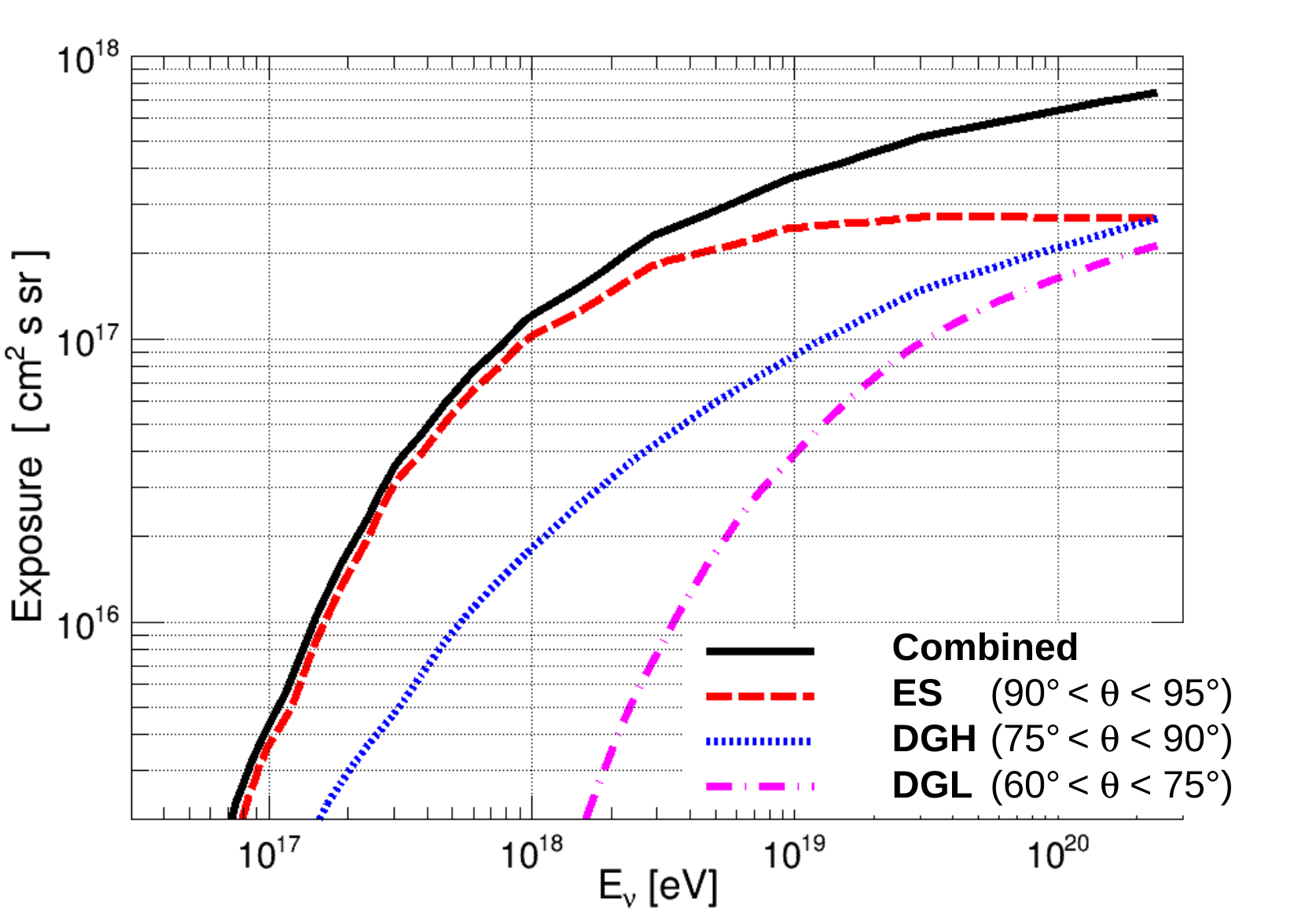}\caption{Total exposure (black solid) and exposures of the ES (dashed red), DGH (dotted blue) and DGL (dot-dashed pink) channels as functions of the neutrino energy, assuming a 1:1:1 flavor ratio at the Earth}\label{fig:exposure}\end{figure}

Given that no candidates have been found, the exposure is calculated in order to obtain flux limits.
The exposure is the neutrino energy dependent integral of the detection probability over surface area, solid angle and time.
For both channels, the detection probability includes the neutrino interaction probability and the probability of the subsequent shower triggering the detector.
For the ES channel, the probability of triggering the detector depends on the probability of the $\tau$ lepton exiting the Earth and decaying close to the detector in one of its decay products that would induce an air shower.
For the DG channels, the probability of triggering the detector depends on the depth of the first interaction in the atmosphere.
The exposures of the different channels and their sum are shown in Figure~\ref{fig:exposure}.
It can be seen that the ES channel dominates the exposure, compensating for the small solid angle and the disadvantage of being sensitive to only one neutrino flavor.
The reason is mainly that the interaction probability of the $\tau$ neutrino is much higher inside the Earth than in the atmosphere as it is proportional to the transversed matter depth.
Also, the ES neutrino induced air showers are extremely well distinguishable from usual UHECR induced air showers at the same inclination because these are so far developed that they are very purely muonic, leaving much narrower traces in the detectors than background showers of lower inclination.
Furthermore, one can see that the exposures of the three channels meet at high energies while the domination of the ES channel is mostly due to lower energies.
This can be explained by the fact that the mean decay length of the $\tau$ lepton is approximately $50~\mathrm{km}\,\frac{E_\tau}{\mathrm{EeV}}$, meaning that at the highest energies it becomes so large that a $\tau$ decay close to the ground becomes less probable.
Note that the mean path length of a $\tau$ lepton in this scenario is not of the order of the size of the Earth since it only skims the Earth, traveling a much shorter path through it.

\begin{figure}[t]
\includegraphics[width=.475\textwidth]{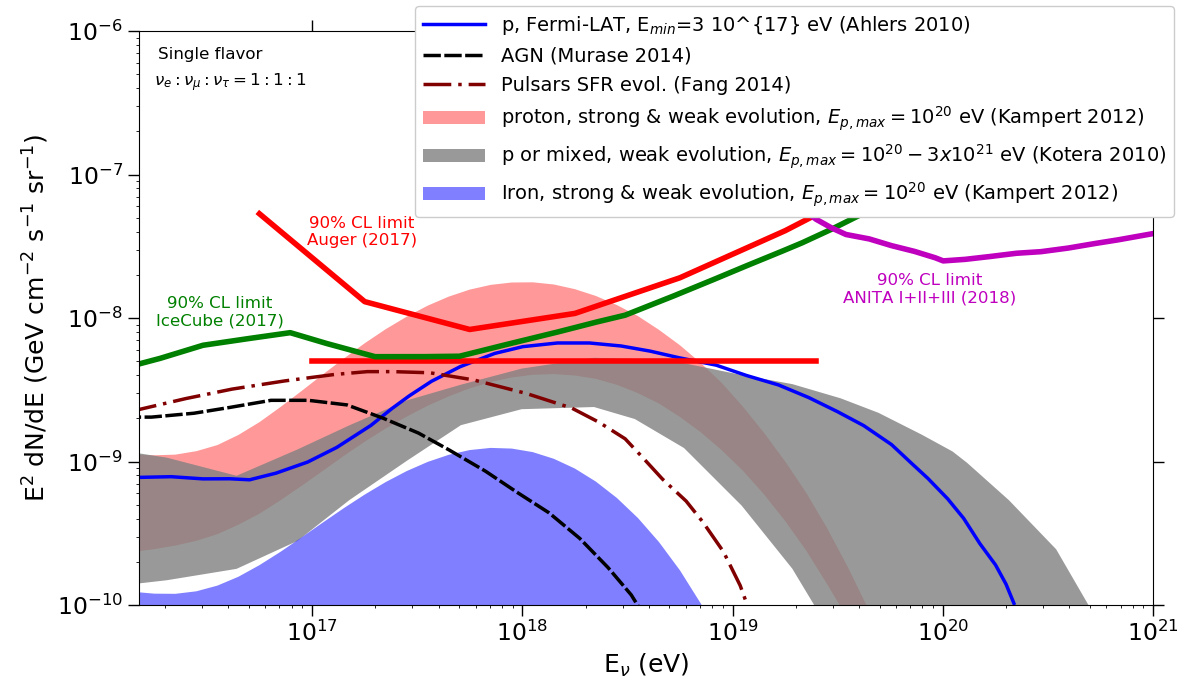}
\caption{Integral upper limit (90\% C.L.) for a diffuse UHE neutrino flux $\mathrm{d}N/\mathrm{d}E_\nu = kE^{-2}$ (lower, longer straight red line), and differential upper limit (upper red line), for a single flavor assuming a 1:1:1 flavor ratio. Corresponding limits from ANITA~\cite{gorham2009new, gorham2010observational, gorham2018constraints} and IceCube~\cite{aartsen2018differential} as well as predictions for several neutrino models (cosmogenic~\cite{kampert2012measurements, kotera2010cosmogenic}, astrophysical~\cite{ahlers2010gzk, murase2014diffuse, fang2014testing}.)}
\label{fig:diffuselim}
\end{figure}

For the calculation of the diffuse limits, a semi-Bayesian extension of the Feldman-Cousins approach is used to account for systematic uncertainties.
The resulting 90\% C.L. upper limits on the event numbers obtained that way correspond to certain flux normalizations, assuming a flux $\propto{}E^{-2}$.
Therefore, they compose the flux limits.
In Figure~\ref{fig:diffuselim}, the limits on a diffuse isotropic flux of single flavor neutrinos are shown.
As can be seen in the plot, the limit calculation has been done in two ways:
On the one hand, an integrated limit is obtained by integrating the flux over the whole energy range of interest and obtaining a flux limit of $5\cdot10^{-9}~\mathrm{GeV}\,\mathrm{cm}^{-2}\,\mathrm{s}^{-1}~E^{-2}$, which is constant in the $E^2$-weighted representation in Figure~\ref{fig:diffuselim}.
The limit is plotted in the energy range of 0.1~EeV through 25~EeV, in which 90\% of the expected sensitivity is contained.
On the other hand, a differential limit is calculated by integrating the flux over energy bins of 0.5 in $\log({E_\nu/\mathrm{EeV}})$.
The limits for these bins are shown in an interpolated representation in Figure~\ref{fig:diffuselim}, revealing that the SD of the Pierre Auger Observatory has its maximal sensitivity close to a neutrino energy of 1~EeV.
Therefore, it is very well suited for detecting cosmological neutrinos resulting from interactions of UHE protons with the Cosmic Microwave Background as indicated by the calculated model fluxes.

In order to obtain confidence levels of exclusion for the models, their associated event rates need to be calculated.
This way, it was found that strong and intermediate source evolution models (with respect to the redshift) with only protons at the sources are mostly excluded with 90\% C.L. by the lack of UHE neutrino candidates due to the models' associated neutrino production~\cite{kampert2012measurements}.
Also, the astrophysical source model in Figure~\ref{fig:diffuselim} assuming active galactic nuclei (AGNs) as the sources of UHECR~\cite{murase2014diffuse} is excluded.

\begin{figure}[t]
\includegraphics[width=.475\textwidth]{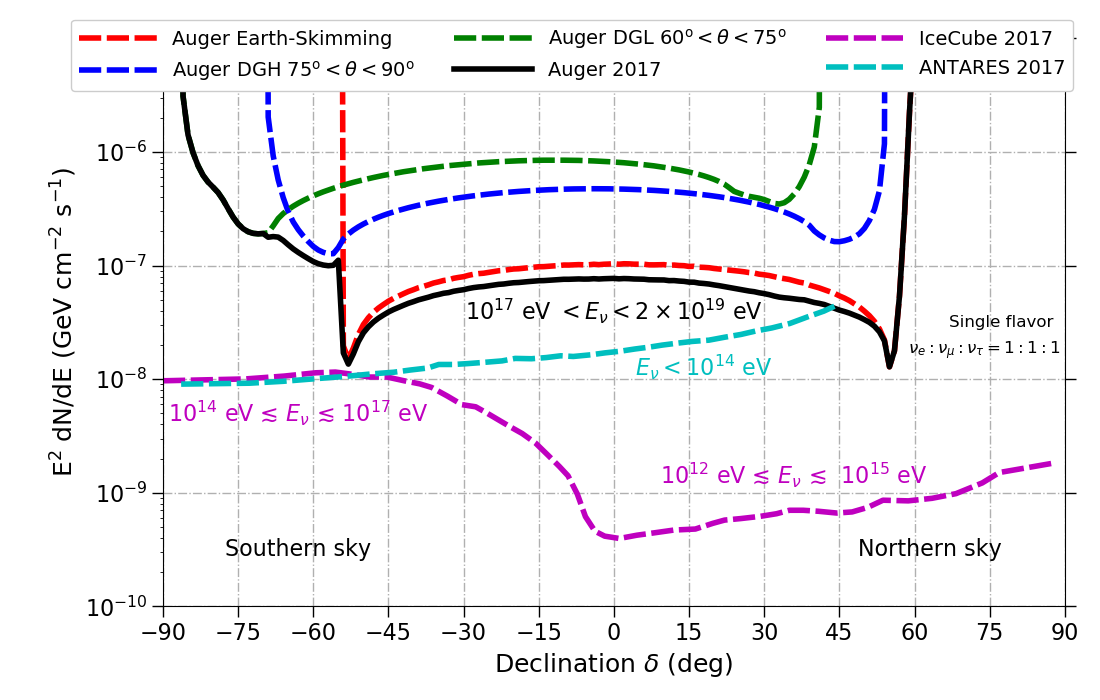}
\caption{Upper limit (90\% C.L). for the flux normalization, assuming flux $\propto{}E^{-2}$, as a function of the source declination for the different neutrino search channels, for a single flavor assuming a 1:1:1 flavor ratio.
Also shown are the limits from IceCube~\cite{aartsen2017all} and ANTARES~\cite{albert2017first}.
Note the different energy ranges.}
\label{fig:PSlim}
\end{figure}

For the calculation of the limits on point-like sources of UHE neutrinos, the direction dependence of the neutrino detection efficiency and the moving field of view (in terms of equatorial coordinates) of the observatory are taken into account.
Since the data for this work has been taken over the course of multiple years, the exposure is uniform in terms of right ascension and the limits on point-like sources depend solely on the declination.
Considering this, the resulting limits are shown in Figure~\ref{fig:PSlim} together with limits from IceCube and ANTARES.
Since the ES channel is dominating the diffuse UHE neutrino exposure, it also dominates the sensitivity to point-like sources in the declination region that it is covering ($-54.5^\circ < \delta < 59.5^\circ$).
The other channels are coming into play for lower declinations, making the combined sensitivity cover the region almost down to the South Celestial Pole.
Considering the complementary energy ranges of the different experiments, it becomes clear that the Pierre Auger Observatory is the only experiment to date that is sensitive to UHE neutrinos from the Northern Hemisphere, in energy ranges multiple orders of magnitude higher than other dedicated neutrino experiments.

\subsection{Sensitivity to transients}
\label{subsec:sensTrans}
\begin{figure}[t]
\includegraphics[width=.475\textwidth]{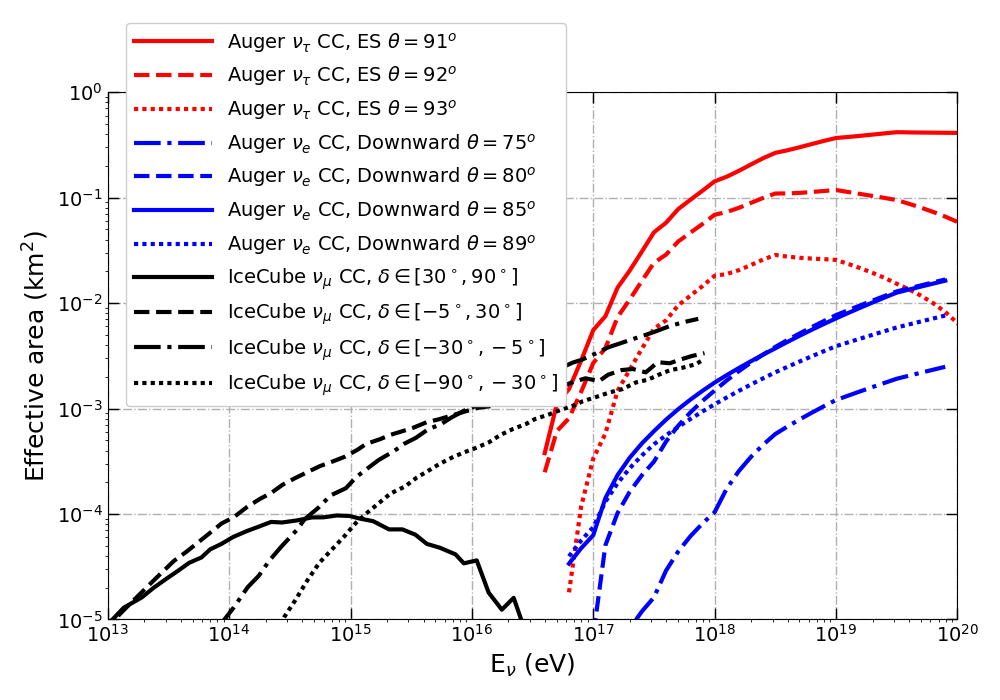}
\caption{Effective area as a function of neutrino energy and for different zenith angles. Red and blue lines: Various zenith angles of ES and DGH neutrino induced showers, respectively. Black lines: Various angular ranges of IceCube.}
\label{fig:Aeff}
\end{figure}

In this subsection, the sensitivity of the SD of the Pierre Auger Observatory to transients is demonstrated, using the effective area $A_\mathrm{eff}$ as an indicator.
The difference to the sensitivity to point-sources or diffuse fluxes is that transients are not steady sources, therefore the sensitivity to them is not only a function of the declination (and energy) but of the efficiency of the detection during the occurrence of the transient.
Thus, it depends on the inclination of the event in the local coordinate system of the observatory.
In Figure~\ref{fig:Aeff}, the effective area as a function of neutrino energy is shown for different zenith angles, along with that of the IceCube Observatory for comparison.
At about $10^{18}~\mathrm{eV}$, the effective area of the Pierre Auger Observatory for neutrinos from slightly below the horizon, and therefore its sensitivity to transients from these directions, is larger than that of IceCube by more than an order of magnitude.
This is particularly interesting because the location of the origin of GW170817, the only known binary neutron star (BNS) merger to date, was slightly below the horizon at the time around the merger as reported in Subsection~\ref{subsec:GW170817}.

\section{Gravitational wave follow-up searches}
\label{sec:GW}
\subsection{Follow-up procedure}
\label{subsec:fuProc}
The Pierre Auger Collaboration and LVC agreed upon the following search parameters for the follow-up of GW in UHE neutrinos:
\begin{itemize}
\item The regular and unmodified neutrino search~\cite{aab2015improved} has to be used.
\item Two search windows were defined, $\pm 500~\mathrm{s}$ around the GW event and the 24-hour period after the GW event.
\item Only times when part of the inner 90\% confidence region of the location of the GW event is in the field of view of the UHE neutrino search of the Pierre Auger Observatory are taken into account.
\end{itemize}
The sensitivity to the different declinations where the event might be located in the sky can be calculated analogously to the sensitivity to point-like sources, using the known effective area depending on the zenith angle, the known geometry of the event, and assuming a conventional flux $\propto{}E^{-2}$.
For the case of the 24-hour period, it is also assumed that the exposure is uniform in right ascension.
Given that no neutrinos have been found, limits can be calculated analogously to the limits on the flux from point-sources.
The properties and results of two of the GW follow-up searches are described in the following.

\subsection{Follow-up of GW events from binary black hole mergers (example: GW151226)}
In this subsection, as an example, the sensitivity to UHE neutrinos in coincidence with a BBH merger is illustrated using GW151216~\cite{abbott2016gw151226}, the second ever reported GW event.
BBH mergers can accelerate cosmic rays to the highest energies provided there are sufficiently strong magnetic fields and a substantial amount of disk debris from the progenitor stars~\cite{kotera2016ultrahigh}.
The fact that very little is known about BBHs is particularly appealing since not many scenarios for them as sources of cosmic rays are excluded so far.

\begin{figure}[t]
\includegraphics[width=.475\textwidth]{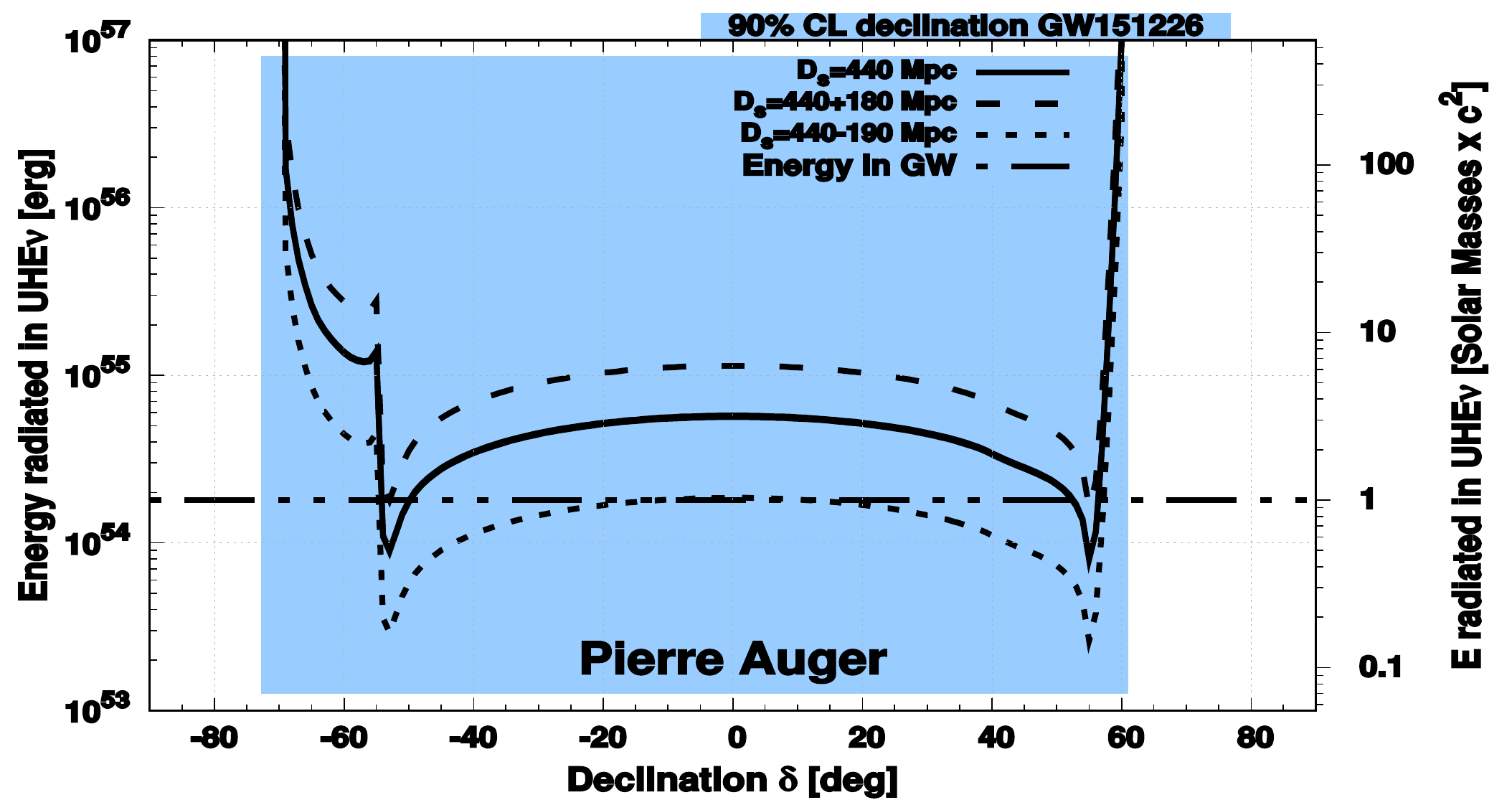}
\caption{Solid: Upper limits (90\% C.L.) on the radiated energy in UHE neutrinos per flavor from the source of GW151226 as a function of the declination~\cite{aab2016ultrahigh}.
Dashed: Analogous upper limits corresponding to reported uncertainties (edges of the central 90\% confidence region) in the distance as indicated.
Dot-dashed (horizontal): Inferred energy radiated in GW~\cite{abbott2016gw151226}.
Filled region: Central 90\% confidence region for the GW event location, projected on the declination.}
\label{fig:GW151226}
\end{figure}

The follow-up search was performed as described in Subsection~\ref{subsec:fuProc}.
During the 24-hour period after the merger, the central 90\% confidence region of the location of the BBH merger was almost completely covered by the field of view of the Pierre Auger Observatory.
This is indicated in Figure~\ref{fig:GW151226}, where the 90\% C.L. declination region is filled in blue and the calculated limit on the energy radiated by the BBH merger in UHE neutrinos is indicated by the solid black line.
As the calculation is analogous to the one of the point-like source limit, the shape of this line resembles the one of the black line in Figure~\ref{fig:PSlim}, indicating the declination-dependent flux limit from point-like sources.
Additionally, the limit on the energy in the form of UHE neutrinos radiated away by the BBH merger is expressed in units of $M_\odot{}c^2$ (with the solar mass $M_\odot$) on the right vertical scale in Figure~\ref{fig:GW151226}.
The energy radiated away in GW is close to $1~M_\odot{}c^2$.
For the declinations with the highest sensitivity, the limit on the energy radiated in the form of UHE neutrinos is lower than this energy by a factor of $\sim 2$.
As a reference, $\lesssim 3\%$ of the energy that is released in the form of GW in BBH mergers would be sufficient to comprise the measured flux of UHECRs if this amount of energy would be used for their acceleration efficiently enough~\cite{kotera2016ultrahigh}.
By searching for UHE neutrinos in coincidence with BBH mergers, this efficiency is indirectly probed.

\subsection{Follow-up of GW170817, a GW event from a binary neutron star merger}
\label{subsec:GW170817}
The event GW170817 originated from the only BNS merger that was directly detected until now.
It triggered an unprecedentedly successful multimessenger campaign, with observations in a big range of the electromagnetic spectrum over multiple weeks~\cite{abbott2017multi}.

A joint neutrino follow-up of this event was performed by the Pierre Auger, IceCube, and ANTARES collaborations~\cite{2017ApJ...850L..35A}.
Among the observations of photons in coincidence with GW170817 was also the detection of a short gamma-ray burst in which high energy photons were emitted.
If these photons are produced in hadronic processes, UHE neutrinos could also be produced~\cite{kimura2017high, fang2017high}.
One difference with respect to GW151226 in terms of GW observation is that VIRGO, a GW detector located in Italy, was operational at the time of GW170817, reducing the size of the 90\% confidence region to about 31 deg$^2$, compared to the $\sim1000~\mathrm{deg}^2$ that were often reported with only the two LIGO detectors.
This improvement substantially reduces the probability of a chance coincidence detection of a UHE neutrino.

\begin{figure}[t]
\includegraphics[width=.475\textwidth]{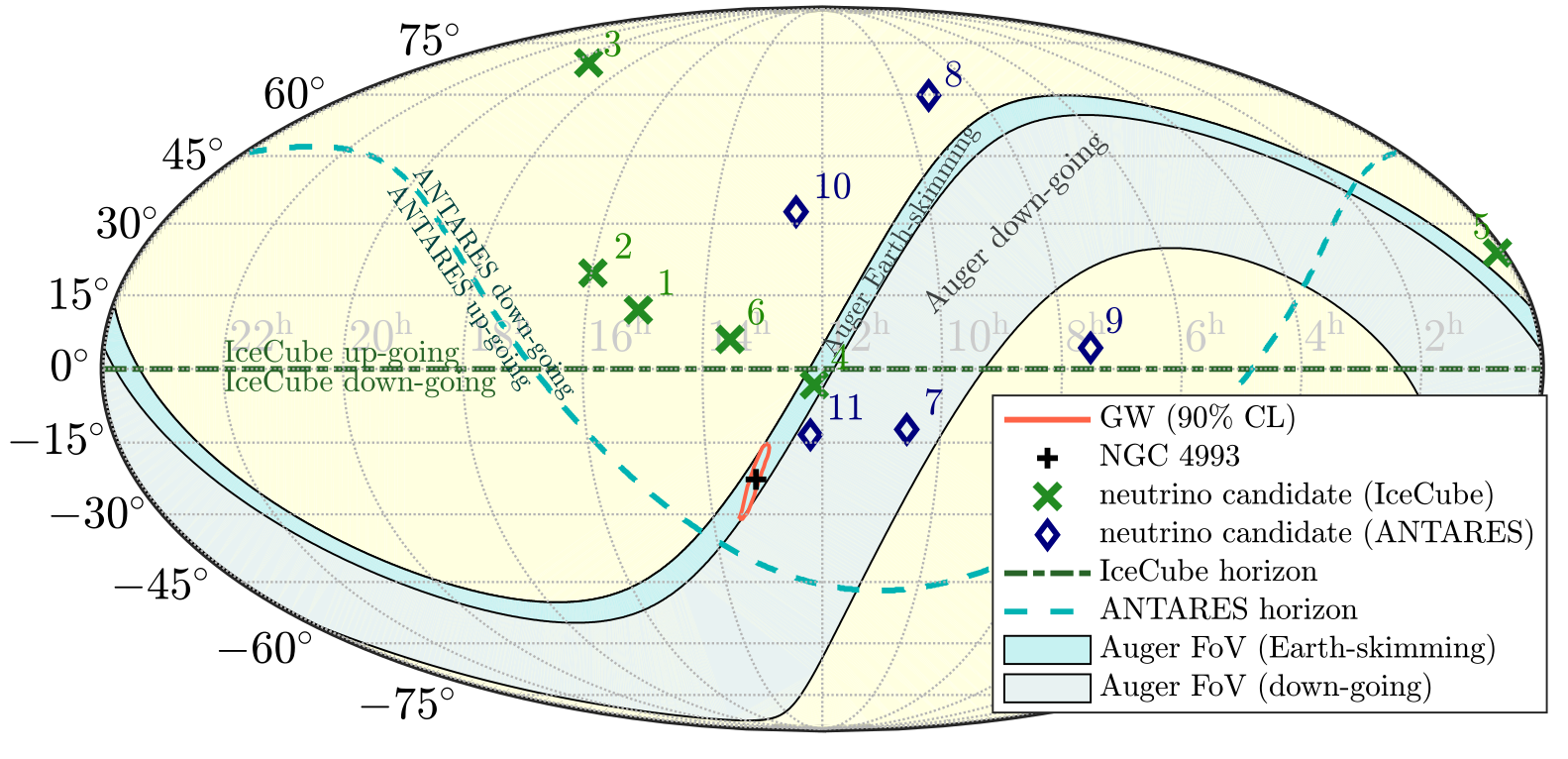}
\caption{Overview of the visibility of GW170817 for IceCube, ANTARES, and the Pierre Auger Observatory.
Red line (black cross): 90\% confidence region of GW170817 (location of its inferred host galaxy, NGC 4993).
Light blue filled bands: Fields of view of the Pierre Auger Observatory, subdivided in ES and DG.}
\label{fig:GW170817_vis}
\end{figure}

\begin{figure}[t]
\includegraphics[width=.475\textwidth]{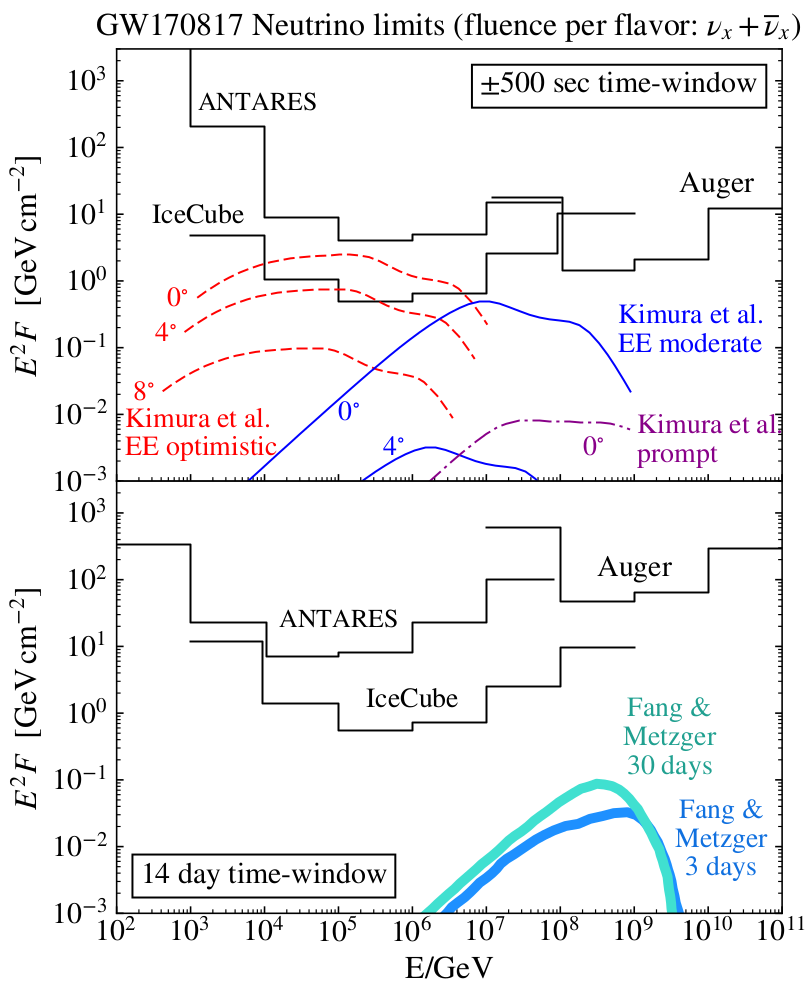}
\caption{Rectangular lines: Limits on the fluence of neutrinos for IceCube, ANTARES, and the Pierre Auger Observatory for the two different time periods.
Smooth lines: Model predictions of neutrino emission for different scenarios and off-axis angles.
}
\label{fig:GW170817_lim}
\end{figure}

The search parameters were agreed upon with IceCube and ANTARES to be modified for this event.
The 1-day time window was extended to 14~days, motivated by the long duration of photon emission from the event's counterpart. 
No associate neutrino candidate has been found by either of the observatories in either of the defined time periods.
Limits on the UHE neutrino flux from GW170817 have been calculated analogously to the previous BBH mergers and have been converted to fluences in order to allow for a comparison between the different experiments.
In Figure~\ref{fig:GW170817_vis}, the visibility of the event at the time of the merger is illustrated.
As mentioned in Subsection~\ref{subsec:sensTrans}, the BNS merger was located slightly below the local horizon at the location of the Pierre Auger Observatory, where the sensitivity to UHE neutrinos is nearly optimal.
This directly translates into a nearly optimal fluence limit for the $\pm 500~\mathrm{s}$ period around the merger as shown in the upper subplot in Figure~\ref{fig:GW170817_lim}.
In this case, the fluence limit above $10^{17}~\mathrm{eV}$ of the Pierre Auger Observatory is one order of magnitude lower than the one of IceCube, making it the most constraining instrument in this energy region.
In contrast to that, for the 14-day period after the merger, the limits from the Pierre Auger Observatory are less competitive since the location of the event is only visible for a limited time every day.
The overall ratio of the sensitivities for this case corresponds approximately to the ratio of the sensitivities to point-like sources at the declination of the GW event ($\delta \approx -23^\circ$) as shown in Figure~\ref{fig:PSlim}, with the difference that here it is presented in terms of energy.
One can see in Figure~\ref{fig:GW170817_vis} that, while IceCube is constraining the flux from the extended emission (EE) optimistic model~\cite{kimura2017high}, the limit set by the Pierre Auger Observatory is the closest to the high-energy cutoff of the moderate emission scenario from~\cite{kimura2017high}.
The lack of neutrino candidates is consistent with an uncommonly low-luminosity GRB or with a typical GRB but observed at a large off-axis angle such that the jets do not point towards the Earth.

\section{Summary and conclusions}
In this work, the searches for UHE neutrinos ($E_\nu > 0.1~\mathrm{EeV}$) with the Pierre Auger Observatory were discussed.
The main observational difference between air showers induced by UHE neutrinos interacting deep in the atmosphere and cosmic ray induced showers at large inclinations is that the former still carry a substantial hadronic and/or electromagnetic component, leaving longer light signals in the water-Cherenkov detectors.
Events induced by such inclined showers are specifically selected and the distributions of the lengths of the light signals in the detectors are used to construct specific variables that allow distinguishing UHE neutrino from cosmic ray induced showers.

Assuming equal neutrino flavor ratios, the sensitivities on the diffuse and point-like source fluxes, using data from 2004-01-01 through 2017-03-31, were shown in Subsection~\ref{subsec:limits}.
No neutrino candidate was found, therefore upper limits on the diffuse flux of UHE neutrinos were calculated.
The Pierre Auger Observatory is a very competitive instrument for UHE neutrino detection, approximately matching the most recent diffuse UHE neutrino limits published by IceCube as shown in Figure~\ref{fig:diffuselim}.
The peak sensitivity at $\sim$1 EeV matches the most common models of cosmological UHE neutrino production.
Using the diffuse limits and the predicted fluxes for these models, it is possible to severely constrain some of the models with a confidence of at least 90\%, in particular those that assume a population of pure proton sources with strong redshift-dependent evolution.
The flux limits for point-like sources shown in Figure~\ref{fig:PSlim} indicate the non-uniformity of the sensitivity in the sky, meaning that there are clearly preferred declinations for UHE neutrino searches with the Pierre Auger Observatory.
In combination, and taking also the energy ranges of the different observatories into account, this means that for some regions in the sky, particularly for the Northern Hemisphere, the Pierre Auger Observatory is the only instrument probing the flux of UHE neutrinos while it is still very competitive in other regions, e.g. around $\delta\approx{}-55^\circ$.
Taking into account that this competitive sensitivity is achieved with a comparably small field of view that is moving across the sky, it also becomes clear that the effective area at high energies must, at least for some directions, be much larger than the one of other observatories as shown in in Figure~\ref{fig:Aeff}.
This large effective area makes the Pierre Auger Observatory a valuable instrument for the observation of transient sources since, given that they are in the observatory's field of view during their occurrence, its time-dependent sensitivity to such sources is much enhanced.
This fact is used in the UHE neutrino follow-up of GW events, which is summarized in the following.

One type of GW follow-up search, which has been performed several times by the Pierre Auger Collaboration, is the follow-up of GW from BBH mergers.
In this work, as an example, the results of the follow-up searches after GW151226 have been presented.
In the 24-hour period after the merger, the range of declinations in which the origin of GW151226 is with 90\% C.L. is almost entirely covered by the UHE neutrino searches with the Auger Observatory.
At the most sensitive declinations, the limit on the energy emitted by the astrophysical counterpart of GW151226 in the form of UHE neutrinos has been found to be lower by a factor of $\sim2$ than the energy emitted in the form of GW as shown in Figure~\ref{fig:GW151226}.

Furthermore, a follow-up search on GW170817 was performed.
Besides the uniqueness of the event as a BNS merger, the counterpart, that was found by photon follow-ups, was located right below the horizon of the Pierre Auger Observatory, where its effective area is maximal, leading to an almost optimal sensitivity to UHE neutrinos.
Consequently, the limits on the neutrino fluence above $10^{17}~\mathrm{eV}$ for the short period of the 1000~seconds surrounding the merger were the best by at least an order of magnitude as can be seen in Figure~\ref{fig:GW170817_lim}.




\bibliographystyle{elsarticle-num}
\bibliography{ref}







\end{document}